\documentclass[3p,twocolumn]{elsarticle}

\usepackage{hyperref}
\usepackage[colorinlistoftodos]{todonotes}

\newcommand{\Agx}{LaCu$_{6-x}$Ag$_{x}$}

\makeatletter
\def\ps@pprintTitle{%
 \let\@oddhead\@empty
 \let\@evenhead\@empty
 \def\@oddfoot{\centerline{}}%
 \let\@evenfoot\@oddfoot}
\makeatother

\begin{document}

\begin{frontmatter}

\title{LaCu$_{6-x}$Ag$_{x}$: A promising host of an elastic quantum critical point }

\author[UTP,QCMD,UMD,NIST]{L. Poudel\corref{cor1}}\ead{lpoudel@vols.utk.edu}
\author[QCMD]{C. de la Cruz}
\author[UTMT]{M. R. Koehler}
\author[MSTD]{M. A. McGuire}
\author[UTMT]{V. Keppens}
\author[UTP,UTMT,MSTD]{D. Mandrus}
\author[QCMD,UTP]{A. D. Christianson}

\cortext[cor1]{Corresponding author}

\address[UTP]{Department of Physics \& Astronomy, University of Tennessee, Knoxville, TN-37996, USA}
\address[QCMD]{Quantum Condensed Matter Division, Oak Ridge National Laboratory, Oak Ridge, TN-37831, USA}
\address[UMD]{Department of Materials Science \& Engineering, University of Maryland, College Park, MD 20742}
\address[NIST]{NIST Center of Neutron Research, Gaithersburg, MD-20899}
\address[UTMT]{Department of Material Science \& Engineering, University of Tennessee, Knoxville, TN-37996, USA}
\address[MSTD]{Materials Science \& Technology Division, Oak Ridge National Laboratory, Oak Ridge, TN-37831, USA
}

\begin{abstract}
Structural properties of \Agx{} have been investigated using neutron and x-ray diffraction, and resonant ultrasound spectroscopy (RUS) measurements. Diffraction measurements indicate a continuous structural transition from orthorhombic ($Pnma$) to monoclinic ($P2_1/c$) structure. RUS measurements show softening of natural frequencies at the structural transition, consistent with the elastic nature of the structural ground state. The structural transition temperatures in \Agx{} decrease with Ag composition until the monoclinic phase is completely suppressed at $x_c$ = 0.225. All of the evidence is consistent with the presence of an elastic quantum critical point in \Agx. 

\end{abstract}


\end{frontmatter}


\begin{figure*}[t]
\centering
\includegraphics[width=\textwidth]{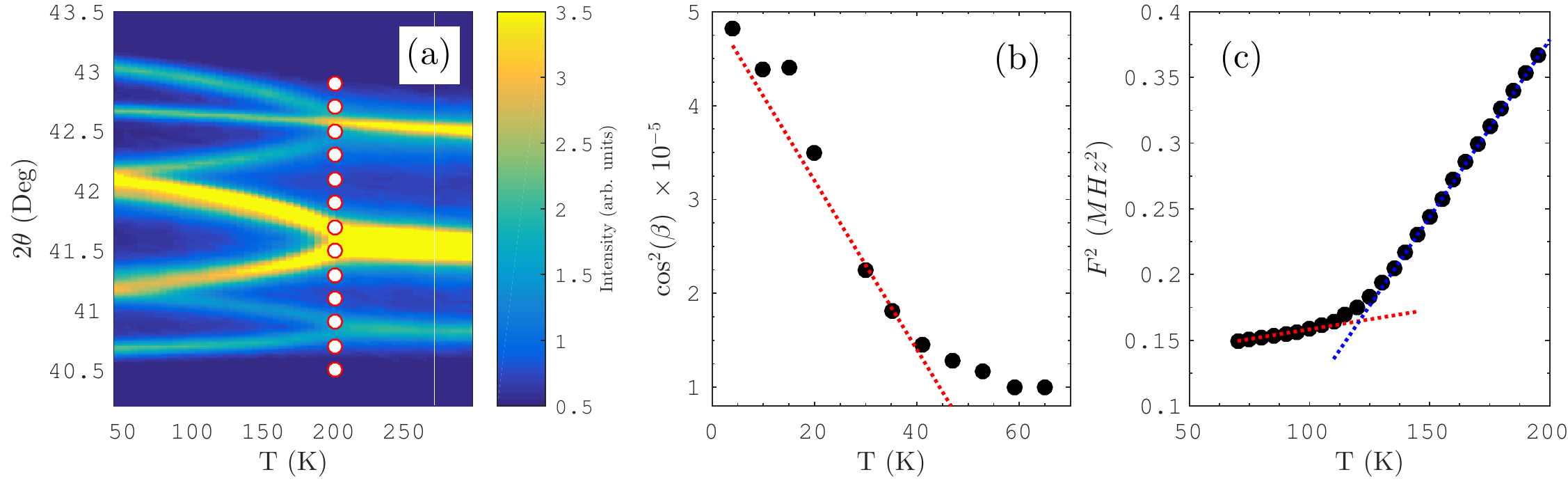}
\caption{Characterization of the structural phase transition in \Agx. (a) Temperature dependence of laboratory x-ray diffraction measurements of $\mathrm{LaCu_{5.875}Ag_{0.125}}$ showing splitting of Bragg peaks. Near $T_S$ (shown by vertical dotted line), due to the monoclinic distortion occurring in the orthorhombic $ab$ plane, the Bragg peaks ($H$ $K$ $L$) with $H\neq0$ and $K\neq 0$ split into a pair of monoclinic peaks. The three peaks in the orthorhombic structure are (122), (220) and (221), which become (22$\pm$1), (20$\pm$2) and (21$\pm$2), respectively. Note the cyclic change in the indexing and lattice parameters between orthorhombic and monoclinic phases. The peak between (21$\pm$2) is the monoclinic (033), which becomes (303) in the orthorhombic phase and overlaps with orthorhombic (221). (b) Temperature dependence of $\cos^2(\beta)$ in $\mathrm{LaCu_{5.8}Ag_{0.2}}$. $T_S$ is obtained by linearly extrapolating $\cos^2(\beta)$ to zero. The red dotted line is a linear extrapolation of the data. (c) RUS measurement of LaCu$_{5.825}$Ag$_{0.175}$ showing the square of a natural frequency ($F^2$) as a function of temperature. The blue and red dotted lines represent the slopes of $F^2$ versus temperature in the orthorhombic and monoclinic phases, respectively. The lines intersect at $T_S$.}
\label{st_transition}
\end{figure*}
 
\section{Introduction}

Quantum criticality continues to be a key pillar of research in condensed matter physics. Prototype examples include magnetic quantum critical points (QCP) in which quantum fluctuations of spins melt an ordered state of matter \cite{sachdev2007quantum,Loh_REV,senthil2004deconfined,coleman2005quantum}. However, in recent years, interest is beginning to shift toward new paradigms of quantum criticality \cite{Rowley2014,hochli_ktio3,ambient_goh,SQCP_IR,scf3}. For example, SrTiO$_3$ and KTaO$_3$ have been identified as being close to a ferroelectric QCP \cite{hochli_ktio3}, and the superconducting state in (Ca,Sr)$_3$Rh$_4$Sn$_{13}$ is associated with the nearby structural QCP \cite{ambient_goh}. Collectively these new examples of quantum criticality provide an opportunity for fresh perspectives and the discovery of new unifying insights. 

Arising out of this emerging interest, an elastic QCP has been theoretically proposed \cite{zacharias}. This type of critical phenomenon is expected to occur when the quantum zero point motion of atoms generates a residual strain in the lattice suppressing the structural ground state. In this sense, the elastic QCP is fundamentally different from the magnetic and other structural counterparts \cite{zacharias}. Furthermore, recent experimental work demonstrates that LaCu$_{6-x}$Au$_x$ shows the promise of hosting an elastic QCP \cite{elastic}. To provide a more comprehensive understanding of the tunability of the structural phase transition leading to an elastic QCP, here we present a study of the related series \Agx{} as a potential candidate of elastic QCP. 

In this paper, we present structural properties of \Agx{} as a function of Ag composition and temperature. The monoclinic phase of \Agx{} is gradually suppressed with Ag substitution. The structural transition is accompanied by a gradual softening of some natural frequencies, as is expected for a continuous elastic phase transition. Linear extrapolation of $T_S$ with $x$ shows that a complete suppression of the monoclinic structure occurs at the critical composition $x_{QCP}$ = 0.225. All of the measurements are consistent with the presence of an elastic QCP in \Agx.

\section{Experimental Details}

Polycrystalline samples of \Agx{} ($x$ = 0, 0.075, 0.1, 0.125, 0.135, 0.15, 0.155, 0.175, 0.2, 0.225, 0.25, 0.3) were synthesized by arc melting the elements La, Cu and Ag in stoichiometric proportions. The phase purity of the sample was characterized by laboratory x-ray measurements at room temperature. Samples with compositions $x$ = 0.075, 0.1, and 0.125 were also measured on a PANalytical X'Pert Pro MPD powder x-ray diffractometer using Cu$K_{\alpha,1}$ radiation ($\lambda~=1.5406~\mbox{\AA}$). For the characterization of the structural transition, diffraction measurements were performed at room temperature and 20 K, and temperature dependence of selected Bragg peaks was obtained in 10 K steps. 

Resonant ultrasound spectroscopy (RUS) measurements were performed on the polycrystalline samples with compositions \Agx, $x$ = 0.135, 0.155, and 0.175 using a set-up as described in the Ref. \cite{migliori2005implementation}. For the measurement, a rectangular parallelepiped sample was held between two transducers. Mechanical resonances within the range 10 - 1000 kHz were collected as a function of temperature.

Neutron diffraction measurements were performed on the samples of \Agx{}($x$ = 0, 0.15, 0.2, 0.225, 0.25) with the HB-2A powder diffractometer at the High Flux Isotope Reactor (HFIR) of Oak Ridge National Laboratory (ORNL). Neutrons of wavelength 1.54 ${\mbox{\AA}}$ were used for the measurement. Collimators containing parallel blades of cadmium coated steel were positioned before the monochromator, sample, and detector with divergence of $12^{\prime}-21^{\prime}-6^{\prime}$ respectively. The sample of mass $\approx$ 5 g was finely ground in a glove box and placed inside a vanadium can with helium as an exchange gas. The can containing the sample was loaded in a closed cycle refrigerator system. Diffraction patterns were collected at room temperature and at 4 K. For $x$ = 0.15 and 0.2, diffraction patterns at several temperatures were obtained near $T_S$. For $x$ = 0.225 and 0.25, diffraction patterns were obtained at room temperature and at 4 K only. The structural parameters were obtained using Rietveld refinement with the FullProf Suite software \cite{rodriguez1990fullprof}.

High resolution x-ray diffraction measurements of LaCu$_{5.7}$Ag$_{0.3}$ were performed using transmission geometry with 11-BM at the Advanced Photon Source at Argonne National Laboratory \cite{wang2008dedicated}. A monochromatic x-ray of wavelength $\lambda = 0.413$ was used for the measurement, which provides a Q-resolution of $\frac{\Delta Q}{Q} = 2\times 10^{-4}$ . The sample was finely ground inside a glove box filled with argon, which was then mixed with amorphous SiO$_2$ in the molar ratio of 1:3 to minimize x-ray absorption. The mixture was packed inside a Kapton tube of 0.8 mm diameter. The Kapton tube containing the sample was spun at 60 Hz to achieve an efficient powder averaging during the measurement.

\begin{table*}[t]
\caption{Structural parameters of LaCu$_{5.7}$Ag$_{0.3}$ obtained from Rietveld refinement of high resolution synchrotron x-ray diffraction measurement (Fig. \ref{diffraction}(a)) at room temperature. The number in parentheses is the error in the last digit. }
\centering
$a$ = 8.20655(5) $\mbox{\AA}$, $b$ = 5.13858(3) $\mbox{\AA}$, $c$ = 10.28100(4) $\mbox{\AA}$\\
$R_P$ = 16.3, $~R_{wp}$ = 18, $~\chi^2$ = 3.1 \\

\begin{tabular}{lllllll}
 \hline
 \hline
 Element\hspace{5pt}		&	Wyck. \hspace{5pt}		&	$x/a$\hspace{5pt}				&	$y/b$ \hspace{5pt}	&	$z/c$	\hspace{5pt} & B \hspace{5pt}			&	Occ.	\\
 \hline
La1 & $4c$ & 0.2598(4) & 0.2500 & 0.4355(3) & 0.82(4) & 1 \\
Cu1 & $8d$ & 0.4374(5) & 0.0002(11) & 0.1895(5) & 0.71(12) & 1 \\
Cu2 & $4c$ & 0.1472(6) & 0.2500 & 0.1397(5) & 0.71(12) & 0.72(3) \\
Ag2 & $4c$ & 0.1472(6) & 0.2500 & 0.1397(5) & 0.46(10) & 0.28(3) \\
Cu3 & $4c$ & 0.8167(7) & 0.2500 & 0.7533(5) & 1.28(13) & 1 \\
Cu4 & $4c$ & 0.5614(8) & 0.2500 & 0.6038(6) & 0.81(12) & 1 \\
Cu5 & $4c$ & 0.9021(7) & 0.2500 & 0.5145(6) & 0.83(7) & 1 \\
\hline
\end{tabular}
\label{table}
\end{table*}

\section{Results and discussion}

\begin{figure}[t]
\centering
\includegraphics[width=0.5\textwidth]{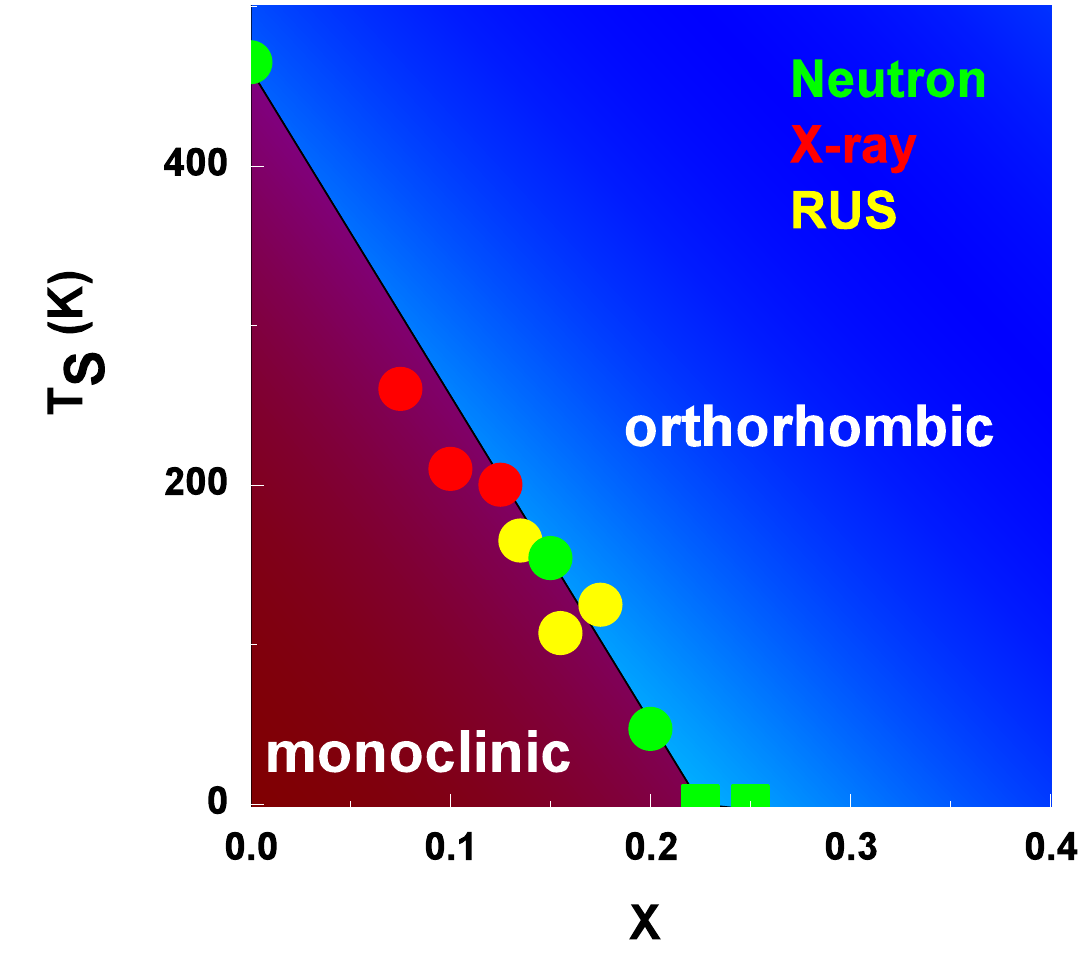}
\caption{Phase diagram of LaCu$_{6-x}$Ag$_x$. $T_S$ decreases linearly with Ag-substitution. No structural transition is observed for $x\geq 0.225$. The squares at T = 0 refer to the observation that no structural transition is observed above 4 K.}
\label{Agphase}
\end{figure}

\begin{figure}[t]
\centering
\includegraphics[width=0.5\textwidth]{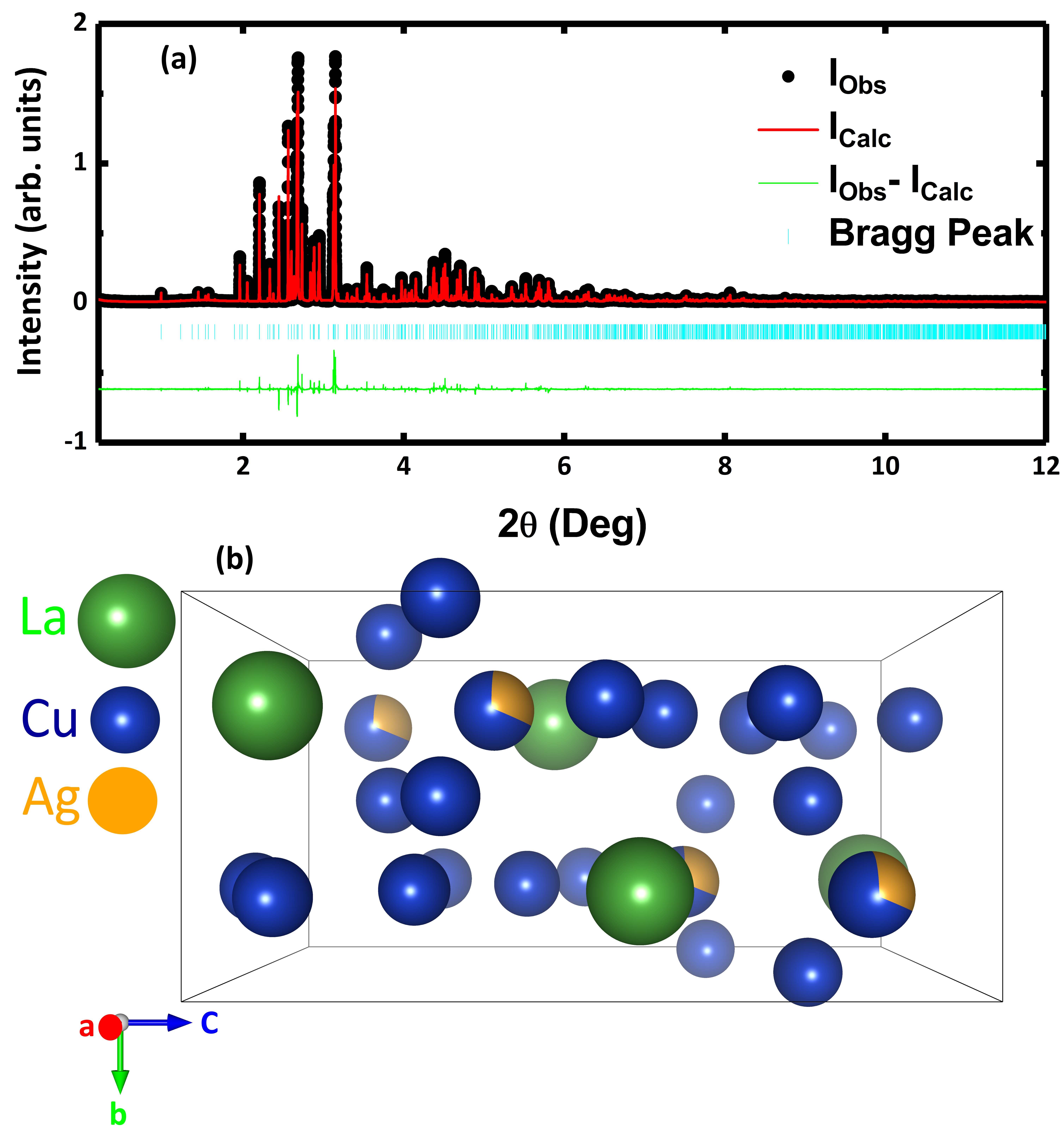}
\caption{(a) Synchrotron x-ray diffraction pattern (black dots) from LaCu$_{5.7}$Ag$_{0.3}$ with the fit (red line) of the orthorhombic crystal structure. The fit was obtained from the Rietveld refinement of the data using the FullProf software. (b) A perspective view of \Agx{} of the orthorhombic structure. The unit cell consists of four formula units of \Agx. Cu atoms are distributed among five different sites. Ag atoms exclusively occupy the Cu2 site. Note: atoms closer to the observer are darker compared to the those further away. }
\label{diffraction}
\end{figure}

Analysis of the room temperature laboratory x-ray diffraction pattern shows that the samples are of high purity and are consistent with either the orthorhombic (space group: $Pnma$) ($x~ \geq ~0.075$) or monoclinic (space group: $P2_1/c$) structure ($x = 0$). The structural phase transition in \Agx{} was characterized by different methods: neutron and x-ray diffraction, and RUS measurements. Using x-ray diffraction, the temperature dependence of selected Bragg peaks was measured. At $T_S$, some of the structural Bragg peaks pertaining to the orthorhombic structure split into two monoclinic peaks, which is shown in Fig. \ref{st_transition}(a). The splitting occurs only for the Bijovet pairs ($H$ $K$ $L$) with $H\neq0$ and $K\neq 0$, which, due to the monoclinic distortion, acquire different $d$-spacing at $T_S$. Neutron diffraction measurements were used to determine the temperature dependence of monoclinic angle $\beta$. Near $T_S$, $\beta$ gradually increases with lowering temperature as expected for a second order phase transition, and consequently, there is a continuous evolution of shear strain ($e_{12} \propto \cos(\beta)$) in the monoclinic phase. Therefore, $\cos^2(\beta)$ is linearly extrapolated to zero for the estimation of $T_S$, as shown in Fig. \ref{st_transition}(b). 

RUS measures the resonances that occur when the frequency of an ultrasonic wave matches with the natural frequency of the sample. The RUS measurements show that some of the natural frequencies gradually become soft as $T_S$ is approached. As the square of a natural frequency ($F^2$) is directly proportional to the combination of elastic constants, we have used $F^2$ as an indirect probe of elastic behavior in \Agx. $T_S$ is characterized by the change in the slope of $F^2$ versus temperature. An example is shown in \ref{st_transition}(c), where the lines representing the slope of $F^2$ in the orthorhombic and monoclinic phases intersect at $T_S$. The change in the slope of $F^2$ versus temperature can be attributed to a complete softening of the $C_{66} = C_{1212}$ elastic constant \cite{Goto1987309,rus_cecu6}, which is expected as the phase transition takes place with softening of corresponding acoustic phonon $\Gamma-X$ and evolution of shear strain $e_{12}$ \cite{elastic}. 

A phase diagram summarizing $T_S$ obtained from diffraction and RUS measurements is presented in Fig. \ref{Agphase}. $T_S$ in \Agx{} linearly decreases with Ag composition. A linear extrapolation of $T_S$ with Ag composition shows that the monoclinic phase is completely suppressed near $x_c$ = 0.225. For the compositions at and above $x_c$, no structural phase transition is observed above 4 K (marked as squares in the phase diagram).

For a detailed understanding of the orthorhombic structure in \Agx{} and in particular, to investigate the distribution of Ag atoms in the orthorhombic unit cell, high resolution synchrotron x-ray measurements were performed for the composition LaCu$_{5.7}$Ag$_{0.3}$. As expected from the phase diagram, the diffraction pattern is consistent with the orthorhombic structure. In addition to the reflections from the main phase, small impurity peaks consistent with elemental copper appear in the diffraction pattern. The intensity of the impurity peaks corresponds to only 0.39\% of copper by weight. No additional impurities were detected. The measured pattern with the fit of orthorhombic crystal structure is shown in Fig. \ref{diffraction}(a). The orthorhombic unit cell consists of four formula units, in which La and Cu/Ag are distributed in one general and five special sites. The crystal structure after the Rietveld refinement is shown in Fig. \ref{diffraction}(b). Details of the Rietveld refinements are presented in Table \ref{table}. 

The result presented here illustrates that the structural properties of \Agx{} are microscopically similar to the related series LaCu$_{6-x}$Au$_x$, which has recently been identified as a host of an elastic QCP, and also to other members of the CeCu$_{6-x}T_x$ family \cite{elastic,poudel,grube1999suppression}. In particular, the Rietveld analysis of the x-ray diffraction measurements shows that the crystal structure of \Agx{} is virtually identical to that of LaCu$_{6-x}$Au$_x$ \cite{elastic,poudel}. The substituent Ag in \Agx{} exclusively occupies the special copper position Cu2, and $T_S$ decreases linearly with chemical substitution as in the case of LaCu$_{6-x}$Au$_x$. Furthermore, RUS measurements indicate that the elastic properties of \Agx{} are similar to the related compound CeCu$_6$, indicating a similarity in structural properties. The structural resemblance of \Agx{} with the LaCu$_{6-x}$Au$_x$ and CeCu$_{6-x}T_x$ family indicates that the suppression of the monoclinic phase in \Agx{} results in an elastic QCP. 

\section{Conclusion}
In conclusion, a structural phase diagram of \Agx{} is constructed using neutron and x-ray diffraction and RUS measurements. High resolution synchrotron x-ray diffraction measurement demonstrates that \Agx{} bears a structural resemblance to the related series LaCu$_{6-x}$Au$_x$ and CeCu$_{6-x}T_x$. The phase transition in \Agx{} is driven by elastic instabilities, with the RUS measurement showing softening of natural frequencies at $T_S$. $T_S$ in \Agx{} can be suppressed with Ag substitution, and the monoclinic phase is completely terminated at the critical composition $x_{QCP}$ = 0.225. The evidence taken together suggests that \Agx{} is a promising host of an elastic QCP. 

\section*{Acknowledgement}
We acknowledge D. Singh for useful discussions and M. Suchomel for assistance with the synchrotron x-ray measurements. The research at the High Flux Isotope Reactor at Oak Ridge National Laboratory is supported by the Scientific User Facilities Division, Office of Basic Energy Sciences, U.S. Department of Energy (DOE). MAM and DM acknowledge support from the U. S. DOE, Office of Science, Basic Energy Sciences, Materials Sciences and Engineering Division. Use of the Advanced Photon Source at Argonne National Laboratory was supported by the U. S. Department of Energy, Office of Science, Office of Basic Energy Sciences, under Contract No. DE-AC02-06CH11357. This manuscript has been authored by UT-Battelle, LLC under Contract No. DE-AC05-00OR22725 with the U.S. Department of Energy. The United States Government retains and the publisher, by accepting the article for publication, acknowledges that the United States Government retains a non-exclusive, paid-up, irrevocable, world-wide license to publish or reproduce the published form of this manuscript, or allow others to do so, for United States Government purposes. The Department of Energy will provide public access to these results of federally sponsored research in accordance with the DOE Public Access Plan (http://energy.gov/downloads/doe-public-access-plan).

 \section*{References}


\begin{thebibliography}{10}
\expandafter\ifx\csname url\endcsname\relax
  \def\url#1{\texttt{#1}}\fi
\expandafter\ifx\csname urlprefix\endcsname\relax\def\urlprefix{URL }\fi
\expandafter\ifx\csname href\endcsname\relax
  \def\href#1#2{#2} \def\path#1{#1}\fi

\bibitem{sachdev2007quantum}
S.~Sachdev, Quantum phase transitions, Wiley Online Library, 2007.

\bibitem{Loh_REV}
H.~v. L\"ohneysen, A.~Rosch, M.~Vojta, P.~W\"olfle, Fermi-liquid instabilities
  at magnetic quantum phase transitions, Rev. Mod. Phys. 79 (2007) 1015--1075.

\bibitem{senthil2004deconfined}
T.~Senthil, A.~Vishwanath, L.~Balents, S.~Sachdev, M.~P. Fisher, Deconfined
  quantum critical points, Science 303~(5663) (2004) 1490--1494.

\bibitem{coleman2005quantum}
P.~Coleman, A.~J. Schofield, Quantum criticality, Nature 433~(7023) (2005)
  226--229.

\bibitem{Rowley2014}
S.~E. Rowley, L.~J. Spalek, R.~P. Smith, M.~P.~M. Dean, M.~Itoh, J.~F. Scott,
  G.~G. Lonzarich, S.~S. Saxena, Ferroelectric quantum criticality, Nat Phys
  10~(5) (2014) 367--372.

\bibitem{hochli_ktio3}
U.~T. H\"ochli, H.~E. Weibel, L.~A. Boatner, Quantum limit of ferroelectric
  phase transitions in $\mathrm{KTa_{1-x}Nb_xO_3}$, Phys. Rev. Lett. 39 (1977)
  1158--1161.

\bibitem{ambient_goh}
S.~K. Goh, D.~A. Tompsett, P.~J. Saines, H.~C. Chang, T.~Matsumoto, M.~Imai,
  K.~Yoshimura, F.~M. Grosche, Ambient pressure structural quantum critical
  point in the phase diagram of $\mathrm{(Sr,Ca)_3Rh_4Sn_{13}}$, Phys. Rev.
  Lett. 114 (2015) 097002.

\bibitem{SQCP_IR}
L.~E. Klintberg, S.~K. Goh, P.~L. Alireza, P.~J. Saines, D.~A. Tompsett, P.~W.
  Logg, J.~Yang, B.~Chen, K.~Yoshimura, F.~M. Grosche, Pressure- and
  composition-induced structural quantum phase transition in the cubic
  superconductor $\mathrm{(Sr,Ca)_3Ir_4Sn_{13}}$, Phys. Rev. Lett. 109 (2012)
  237008.

\bibitem{scf3}
S.~U. Handunkanda, E.~B. Curry, V.~Voronov, A.~H. Said, G.~G. Guzm\'an-Verri,
  R.~T. Brierley, P.~B. Littlewood, J.~N. Hancock, Large isotropic negative
  thermal expansion above a structural quantum phase transition, Phys. Rev. B
  92 (2015) 134101.

\bibitem{zacharias}
M.~Zacharias, I.~Paul, M.~Garst, Quantum critical elasticity, Phys. Rev. Lett.
  115 (2015) 025703.

\bibitem{elastic}
L.~Poudel, A.~F. May, M.~R. Koehler, M.~A. McGuire, S.~Mukhopadhyay, S.~Calder,
  R.~E. Baumbach, R.~Mukherjee, D.~Sapkota, C.~de~la Cruz, D.~J. Singh,
  D.~Mandrus, A.~D. Christianson, Candidate elastic quantum critical point in
  $\mathrm{LaCu_{6-x}Au_x}$, Phys. Rev. Lett. 117 (2016) 235701.

\bibitem{migliori2005implementation}
A.~Migliori, J.~Maynard, Implementation of a modern resonant ultrasound
  spectroscopy system for the measurement of the elastic moduli of small solid
  specimens, Review of Scientific Instruments 76~(12) (2005) 121301.

\bibitem{rodriguez1990fullprof}
J.~Rodr{\'i}guez-Carvajal, Recent advances in magnetic structure determination
  by neutron powder diffraction, Physica B: Condensed Matter 192~(1-2) (1993)
  55 -- 69.

\bibitem{wang2008dedicated}
J.~Wang, B.~H. Toby, P.~L. Lee, L.~Ribaud, S.~M. Antao, C.~Kurtz,
  M.~Ramanathan, R.~B. Von~Dreele, M.~A. Beno, A dedicated powder diffraction
  beamline at the advanced photon source: Commissioning and early operational
  results, Review of Scientific Instruments 79~(8) (2008) 085105.

\bibitem{Goto1987309}
T.~Goto, T.~Suzuki, A.~Tamaki, T.~Fujimura, Y.~{\=O}nuki, T.~Komatsubara,
  Elastic properties of the $\mathrm{Kondo}$ lattice compound
  $\mathrm{CeCu_6}$, J. Magn. Magn. Mater. 63--64 (1987) 309--311.

\bibitem{rus_cecu6}
T.~Suzuki, T.~Goto, A.~Tamaki, T.~Fujimura, Y.~\={O}nuki, T.~Komatsubara,
  Elastic soft mode and crystalline field effect of $\mathrm{Kondo}$ lattice
  substance: $\mathrm{CeCu_{6}}$, J. Phys. Soc. Jpn. 54~(7) (1985) 2367--2370.

\bibitem{poudel}
L.~Poudel, C.~de~la Cruz, E.~A. Payzant, A.~F. May, M.~Koehler, V.~O. Garlea,
  A.~E. Taylor, D.~S. Parker, H.~B. Cao, M.~A. McGuire, W.~Tian, M.~Matsuda,
  H.~Jeen, H.~N. Lee, T.~Hong, S.~Calder, H.~D. Zhou, M.~D. Lumsden,
  V.~Keppens, D.~Mandrus, A.~D. Christianson, Structural and magnetic phase
  transitions in $\mathrm{CeCu_{6-x}T_x}$ ($\mbox{T}$ = $\mathrm{Ag}$,
  $\mathrm{Pd}$), Phys. Rev. B 92 (2015) 214421.

\bibitem{grube1999suppression}
K.~Grube, W.~H. Fietz, U.~Tutsch, O.~Stockert, H.~v. L\"ohneysen, Suppression
  of the structural phase transition in $\mathrm{CeCu_{6-x}Au_x}$ by pressure
  and $\mathrm{Au}$ doping, Phys. Rev. B 60 (1999) 11947--11953.

\end{thebibliography}

\end{document}